\documentclass{article}
\usepackage{a4wide}
\usepackage{latexsym}
\usepackage{amsmath}
\usepackage{amssymb}
\usepackage{ifthen}
\usepackage{mathrsfs}
\usepackage{authblk}

\usepackage{amsthm}

\usepackage{stmaryrd}

\usepackage{pstricks}
\usepackage{pst-node}
\usepackage{pst-tree}
\usepackage{pst-text}

\psset{dotsize=2mm,linewidth=0.7pt}
\newpsobject{showgrid}{psgrid}{subgriddiv=1,griddots=10,gridlabels=6pt}

\newcommand{\nc}{\newcommand}
\nc{\rnc}{\renewcommand} \nc{\nev}{\newenvironment}
\rnc{\subsection}{\secdef\ssa\ssb}
\nc{\ssa}[2][default]{\par\vspace{1ex}\refstepcounter{subsection}\noindent\textbf{\thesubsection.
#1. }} \nc{\ssb}[1]{\par\vspace{2ex}\noindent\textbf{#1. }}

\rnc{\subsubsection}{\secdef\sssa\sssb}
\nc{\sssa}[2][default]{\par\vspace{1ex}\refstepcounter{subsubsection}\noindent\textit{\thesubsubsection.
#1. }} \nc{\sssb}[1]{\par\vspace{1ex}\noindent\textit{#1. }}

\makeatletter \rnc{\@seccntformat}[1]{{\normalfont\bfseries{\csname
the#1\endcsname}\hspace{1pt}.\hspace{0.4em}}}
\rnc{\section}{\@startsection
        {section}%
        {1}%
        {0mm}%
        {-\baselineskip}%
        {0.5\baselineskip}%
        {\normalfont\normalsize\bfseries\centering}%
}
\renewcommand{\@makecaption}[2]{\begin{center}#1. #2\end{center}}
\makeatother

\newtheorem{theo}{Theorem}[section]
\newtheorem{lem}[theo]{Lemma}

\newtheorem{prop}[theo]{Proposition}

\theoremstyle{definition}
\newtheorem{defn}[theo]{Definition}
\newtheorem{rem}[theo]{Remark}

\rnc{\proof}{\smallskip\noindent\textit{Proof: }}
\nc{\proofend}{\hfill$\Box$\vspace{\topsep}\par}

\rnc{\labelenumi}{(\arabic{enumi})} \rnc{\labelitemi}{\text{--}}
\rnc{\phi}{\varphi} \rnc{\epsilon}{\varepsilon}
\nc{\bigmid}{\;\big|\;} \nc{\Bigmid}{\;\Big|\;}
\rnc{\max}{\textup{max}} \rnc{\min}{\textup{min}}
\rnc{\log}{\textup{log}\;}

\newlength{\probwidth}
\nc{\prob}[3][9]{
\begin{center}
  \normalfont\fbox{
   \begin{tabular}[t]{
     rp{#1cm}}\textit{Instance:}&#2. \\
     \textit{Problem:}&#3
   \end{tabular}}
\end{center}}

\nc{\pprob}[4][9]{
\begin{center}
   \normalfont\fbox{
    \begin{tabular}[t]{
     rp{#1cm}}\textit{Instance:}&#2. \\
     \textit{Parameter:}&#3. \\
     \textit{Problem:}&#4
   \end{tabular}}
\end{center}}

\nc{\nprob}[4][9]{
\begin{center}
  \normalfont\fbox{

\addtolength{\probwidth}{#1cm}\parbox{\probwidth}{\textsc{#2}\\\hspace*{1.5em}
     \begin{tabular}[t]{
      rp{#1cm}}\textit{Instance:}&#3. \\
      \textit{Problem:}&#4
     \end{tabular}}}
\end{center}}

\nc{\npprob}[5][9]{
\begin{center}
  \normalfont\fbox{

\addtolength{\probwidth}{#1cm}\parbox{\probwidth}{\textsc{#2}\\\hspace*{1.5em}
    \begin{tabular}[t]{
     rp{#1cm}}\textit{Instance:}&#3. \\
     \textit{Parameter:}&#4. \\
     \textit{Problem:}&#5
    \end{tabular}}}
\end{center}}

\nc{\nppxrob}[5][9]{ \normalfont\fbox{

\addtolength{\probwidth}{#1cm}\parbox{\probwidth}{\textsc{#2}\\\hspace*{1.5em}
   \begin{tabular}[t]{
    rp{#1cm}}\textit{Instance:}&#3. \\
    \textit{Parameter:}&#4. \\
    \textit{Problem:}&#5
   \end{tabular}}}}

\nc{\nppprob}[5][4]{
\begin{center}
  \normalfont\fbox{

\addtolength{\probwidth}{#1cm}\parbox{\probwidth}{\textsc{#2}\\\hspace*{1.5em}
    \begin{tabular}[t]{
     rp{#1cm}}\textit{Instance:}&#3. \\
     \textit{Parameter:}&#4. \\
     \textit{Problem:}&#5
    \end{tabular}}}
\end{center}}

\nc{\FOR}{\textbf{for}} \nc{\FORALL}{\textbf{for all}}
\nc{\TO}{\textbf{to}} \nc{\DO}{\textbf{do}} \nc{\OD}{\textbf{od}}
\nc{\IF}{\textbf{if}} \nc{\FI}{\textbf{fi}}
\nc{\THEN}{\textbf{then}} \nc{\ELSE}{\textbf{else}}
\nc{\WHILE}{\textbf{while}} \nc{\REPEAT}{\textbf{repeat}}
\nc{\UNTIL}{\textbf{until}} \nc{\OR}{\textbf{or}}
\nc{\AND}{\textbf{and}} \nc{\PRINT}{\textbf{print}}

\nc{\im}[1]{\item\hspace{#1cm}}
\nev{algorithm}{\begin{enumerate}\rnc{\labelenumi}{\textit{\small
\arabic{enumi}.}}\rnc{\itemsep}{0ex}}{\end{enumerate}}

\nc{\fpcl}[1]{\left[#1\right]_{\text{\upshape fp}}}
\nc{\pr}{\le^{\text{\normalfont fp}}_m} \nc{\FPT}{\textup{FPT}}
\nc{\EPT}{\textup{EPT}} \nc{\SUBEPT}{\textup{SUBEPT}}

\nc{\fpt}{\textup{fpt}} \nc{\fptT}{\textup{fpt-T}}
\nc{\W}[1]{\text{$\textup{W}[#1]$}}
\nc{\M}[1]{\text{$\textup{M}[#1]$}}
\nc{\MS}[2]{\text{$\textup{M}^{#1}[#2]$}} \nc{\MINI}[1]{\mbox{\small
\rm MINI[$#1$]}} \nc{\WP}{\textup{W[P]}} \nc{\AWP}{\textup{AW[P]}}

\rnc{\S}[1]{\text{$\textup{S}[#1]$}} \nc{\SP}{\textup{S[P]}}
\nc{\MP}{\textup{M[P]}}

\nc{\PTIME}{\textup{PTIME}} \nc{\APTIME}{\textup{APTIME}}
\nc{\PSPACE}{\textup{PSPACE}} \nc{\NP}{\textup{NP}}

\nc{\DTIME}[1]{\textup{DTIME}(#1)}

\nc{\PNPTC}{\mbox{$\textup{P}[{\textsc{tc}}]\ne
\textup{NP}[{\textsc{tc}}]$}} \nc{\str}[1]{\ensuremath{\mathcal #1}}
\nc{\cls}[1]{\ensuremath{\mathbf #1}}

\nc{\ceil}[1]{\left\lceil#1\right\rceil}
\nc{\floor}[1]{\left\lfloor#1\right\rfloor}

\nc{\bende}{\eqno$\Box$} \nc{\benda}{\tag*{$\Box$}}

\nc{\pa}{\kappa}

\nc{\MSO}{\textup{MSO}} \nc{\EMSO}{\textup{EMSO}}

\nc{\rand}[1]{\marginpar{\raggedright\footnotesize #1}}
\nc{\brand}[1]{\rand{\textbf{B: }#1}}
\nc{\yrand}[1]{\rand{\textbf{Y: }#1}}

\nc{\ma}[1]{\mathbb #1}

\nc{\Pt}{\textup{P}}

\nc{\pEIS}{\ensuremath{p\textsc{-Edge-Induced-Subgraph}}}

\nc{\EIS}{\textsc{Edge-Induced-Subgraph}}

\nc{\pCEIS}{\ensuremath{p\textsc{-\#Edge-Induced-Subgraph}}}

\nc{\ltw}{\textup{ltw}}
\nc{\tw}{\textup{tw}}

\nc{\pMCLTWFO}[1]{\ensuremath{p\textsc{-Mc-Ltw}_{#1}\textsc{-FO}}}

\nc{\pSCLTW}[1]{\ensuremath{p\textsc{-Scattered-Set-Ltw}_{#1}}}

\nc{\pSCTW}{\ensuremath{p\textsc{-Scattered-Set-Tw}}}

\nc{\FO}{\textup{FO}}

\title{\Large\bf The parameterized complexity of $k$-edge induced subgraphs}
\date{}

\author{\normalsize Bingkai Lin}
\author{\normalsize Yijia Chen}
\affil{\normalsize Department of Computer Science and Engineering \\
\normalsize Shanghai Jiaotong University \\}

\begin{document}

\maketitle

\begin{abstract}
We prove that finding a $k$-edge induced subgraph is fixed-parameter
tractable, thereby answering an open problem of Leizhen
Cai~\cite{hancai}. Our algorithm is based on several combinatorial
observations, Gauss' famous \emph{Eureka} theorem~\cite{and}, and a
generalization of the well-known \fpt-algorithm for the
model-checking problem for first-order logic on graphs with locally
bounded tree-width due to Frick and Grohe~\cite{frigro}. On the
other hand, we show that two natural counting versions of the
problem are hard. Hence, the $k$-edge induced subgraph problem is
one of the very few known examples in parameterized complexity that
are easy for decision while hard for counting.
\end{abstract}

\section{Introduction}
Induced subgraphs are one of the most natural substructures in
graphs. They capture many different combinatorial objects, e.g.,
clique, independent set, chordless path. Thus, a great number of
algorithmic problems are about finding certain induced subgraphs,
and their complexity is among the mostly extensively studied in
algorithmic graph
theory~\cite{cai96,cheturwey,chukawsey,hashoff,khoram,mossik,mosthi,papyan}.
Induced subgraphs with distinct number of edges have also been studied in graph theory~\cite{AlonB89,AlonK09}. In this paper, we are mainly
interested in the problem of finding an induced subgraph which
contains exactly $k$ edges, i.e., a $k$-edge induced subgraph. This
problem is equivalent to solving a special $0\text{-}1$ quadratic
Diophantine  equation $x^TAx=k$, where $A$ is the adjacent matrix of
$G$, $x\in\{0,1\}^n,n=|V(G)|$.

It is not difficult to prove that the $k$-edge induced subgraph
problem is \NP-hard by a reduction from the clique problem. So we approach
the problem via parameterized complexity~\cite{dowfel,flugro,nie}
and treat $k$ as the parameter:
\npprob[8]{\pEIS}{A graph $G$ and $k\in \mathbb N$}{$k$}{Decide
whether $G$ contains a $k$-edge induced subgraph.}

As the main result of our paper, we show that $\pEIS$ is
fixed-parameter tractable. In fact, there are special cases of
\pEIS\ whose fixed-parameter tractability has been known for a
while. Since we can define a $k$-edge induced subgraph by a
first-order sentence, using logic machinery, it can be shown that
$\pEIS$ is fixed-parameter tractable if the graph $G$ has bounded
tree-width~\cite{cou}, bounded local tree-width~\cite{frigro}, etc.,
or most generally locally bounded expansion~\cite{dvokratho}.
Unfortunately, the class of all graphs containing a $k$-edge induced
subgraph does not possess any of these bounded measures. As another
previously known case, using his \emph{Random Separation}
method~\cite{caicha} and Ramsey's Theorem, Cai~\cite{cai} gave a
very nice combinatorial algorithm that solves \pEIS\ when the
parameter $k$ is a \emph{triangular number}, i.e., $k=\binom{m}{2}$
for some $m\in \mathbb N$. However, it looks very difficult to adapt
Cai's algorithm to handle arbitrary $k$. Therefore neither logic nor
combinatorial approach so far seems to be sufficient to settle the
complexity of \pEIS\ by its own. So our \fpt-algorithm is a rather
tricky combination of these two methods.

\subsection*{Our approach}
As just mentioned, our starting pointing is that the existence of a
$k$-edge induced subgraph can be characterized by a sentence of
first-order logic (\FO) which depends on $k$ only. It is a
well-known result of Frick and Grohe~\cite{frigro} that the
model-checking problem for \FO\ on graphs of bounded \emph{local
tree-width} is fixed-parameter tractable. The local tree-width for a
graph is a function bounding the tree-width of the induced subgraphs
on the neighborhoods within a certain radius of every vertex. For
instance, bounded-degree graphs have bounded local tree-width. These
give immediately the fixed-parameter tractability of \pEIS\ on
graphs with bounded degree\footnotemark. \footnotetext{This is also
a direct consequence of Seese's result that the model-checking
problem for \FO\ on bounded-degree graphs is fixed-parameter
tractable~\cite{see}. But we find it more natural to work with
bounded local tree-with in the following generalization.}

With some more efforts, the above result can be extended to graphs
$G$ with degree bounded by a function of the parameter $k$. In that
case, we can say the degree $\deg(v)$ of each vertex $v$ is
sufficiently small. The corresponding fpt-algorithm generalizes
Frick and Grohe's Theorem to graphs with local tree-width bounded by
a function of both the radius of the neighborhoods and an additional
parameter. As a dual, if $\deg(v)$ of each vertex $v$ in $G$ is
sufficiently large, or more precisely, the complement of $G$ has
degree bounded by a function of $k$, then we can decide $\pEIS$ in
\fpt\ time, too.

Moving one step further, we consider graphs in which each $\deg(v)$
is either sufficiently small or sufficiently large, e.g., an
$n$-star. We call such graphs \emph{degree-extreme}. Using the same
logic machinery as above, we then are able to show the
fixed-parameter tractability of $\pEIS$ on degree-extreme graphs.

\medskip
Assume that the graph $G$ is not degree-extreme, i.e., there exists
a vertex $v_0$ whose degree is neither sufficiently small nor
sufficiently large. We partition the vertex set of $G$ into two sets
$V_1$ and $V_2$, where $V_1$ contains all vertices adjacent to $v_0$
and $V_2$ the remaining vertices. Then both $V_1$ and $V_2$ are
relatively large. Note possibly there are many edges between $V_1$
and $V_2$. Nevertheless, we can compute a vertex set $B$ in $G$ such
that every edge between $V_1$ and $V_2$ has one vertex in $B$; and
if $B$ is large enough, we can show that $G$ contains a $k$-edge
induced subgraph. Otherwise, the graph $G$ consists of two induced
subgraphs $G[V_1]$ and $G[V_2]$, plus the edges between $V_1$ and
$V_2$ adjacent to the set $B$ of bounded size. In case $G[V_1]$ and
$G[V_2]$ are both degree-extreme, we call such a graph $G$ a
\emph{bridge} (of two degree-extreme graphs). By the logic method
again, we prove that $\pEIS$ is fixed-parameter tractable on
bridges.

Now we are left with the case that at least one of $G[V_1]$ and
$G[V_2]$ is not degree-extreme, say $G[V_1]$. Then we repeat the
above procedure on $G[V_1]$ to get a partition $V_{11}\mid V_{12}$
of $V_1$. And again, both $V_{11}$ and $V_{12}$ are sufficiently
large. Arguing as before, either we already know $G[V_1]$, and hence
$G$, contains a $k$-edge induced subgraph, or there is a set $B_1$
of bounded size such that every edge between $V_{11}$ and $V_{22}$
intersects $B_1$.

Finally we remove the vertex set $B_0:= B\cup B_1$ from $G$. Then
$G[V\setminus B_0]$ is the disjoint union of $G[V_{11}\setminus
B_0]$, $G[V_{12}\setminus B_0]$ and $G[V_2\setminus B_0]$. Moreover,
all three induced subgraphs are so large that, by Ramsey's Theorem,
either one of them contains a large independent set, or we have
three large disjoint cliques which are not adjacent to each other.
For both cases, we show that $G[V\setminus B_0]$, and hence $G$,
contains a $k$-edge induced subgraph. As a matter of fact, the
second case is an easy consequence of a famous number-theoretic
result of Gauss which states that \emph{every natural number is the
sum of three triangular numbers.}

\medskip
We should mention that the running time of our algorithm in terms of
the parameter $k$ is astronomical, \emph{triple exponential} at
least. But we hope that similar as it happened in many other cases
the knowledge that the k-edge problem is fixed-parameter tractable
will encourage to look for faster algorithms or at least for
algorithms useful in practice for concrete classes of instances of
the problem.


\subsection*{Counting $k$-edge induced subgraphs}
We also study the parameterized complexity of computing the number
of $k$-edge induced subgraphs. For most natural problems, if the
decision version is easy, then so is the counting problem. However,
it turns out that two natural counting versions of $\pEIS$ are both
hard. To the best of our knowledge, there are only very few natural
problems which exhibit such a phenomenon~\cite{flugro04,cheflu07}.

\medskip
\subsection*{Organization of our paper}
In Section~\ref{sec:pre} we introduce necessary background and fix
our notations. We prove all required combinatorial results in
Section~\ref{sec:comb}. In particular, we present several simple
structures in a graph which, if exist, guarantee the existence of a
$k$-edge induced subgraph. Then in Section~\ref{sec:logic} we
establish the fixed-parameter tractability of $\pEIS$ on
degree-extreme graphs and bridges using model-checking problems for
\FO. We present our \fpt-algorithm for \pEIS\ by putting all the
pieces together in Section~\ref{sec:algorithm}. Finally in
Section~\ref{sec:count} we prove the hardness of the counting
problems. For readers not familiar with~\cite{frigro}, we provide a
proof of the easy generalization of Frick and Grohe's algorithm in
an appendix.

\section{Preliminaries}\label{sec:pre}
$\mathbb N$ and $\mathbb N^+$ denote the sets of natural numbers
(that is, nonnegative integers) and positive integers, respectively.
For a natural number $n$ let $[n] := \{1, \ldots, n \}$.

We denote the alphabet $\{0,1\}$ by $\Sigma$ and identify problems
with subsets $Q$ of $\Sigma^*$. Clearly, as done mostly, we present
concrete problems in a verbal, hence uncodified form over $\Sigma$.

For every set $S$ we use $|S|$ to denote its size. Moreover we let
$\binom{S}{2}$ be the set of all two-element subsets of $S$,
i.e.,$\big\{\{a,b\} \bigmid \text{$a,b\in S$ and $a\ne b$}\big\}$. A
triangular number is $\binom{k}{2} := \big|\binom{[k]}{2} \big|$ for
some $k\in \mathbb N$. In particular, $\binom{0}{2}=\binom{1}{2}=0$.

\subsection*{Parameterized complexity}
A \emph{parameterized problem} is a pair $(Q,\kappa)$ consisting of
a classical problem $Q \subseteq \Sigma^*$ and a polynomial time
computable {\em parameterization} $\kappa : \Sigma^* \to \mathbb N$.

An algorithm $\mathbb A$ is an \emph{$\fpt$-algorithm with respect
to a parameterization $\kappa$} if for every $x\in \Sigma^*$ the
running time of $\mathbb A$ on $x$ is bounded by $f(\kappa(x))\cdot
|x|^{O(1)}$ for a computable function $f:\mathbb N\to \mathbb N$. Or
equivalently, we say that the algorithm $\mathbb A$ runs in \fpt\
time. A parameterized problem $(Q,\kappa)$ is \emph{fixed-parameter
tractable} if there is an \fpt-algorithm with respect to $\kappa$
that decides $Q$.

Let $(Q,\kappa)$ and $(Q',\kappa')$ be two parameterized problems.
An \emph{\fpt-reduction} from $(Q,\kappa)$ to $(Q',\kappa')$ is a
mapping $R:\Sigma^*\to \Sigma^*$ such that:
\begin{itemize}
\item For every $x\in \Sigma^*$ we have $x\in Q$ if and only if
    $R(x)\in Q'$.

\item $R$ is computable by an \fpt-algorithm.

\item There is a computable function $g:\mathbb N\to \mathbb N$
    such that $\kappa'(R(x)) \le g(\kappa(x))$ for all $x\in
    \Sigma^*$.
\end{itemize}
It is easy to see that if there is an \fpt-reduction from
$(Q,\kappa)$ to $(Q',\kappa')$, and if $(Q',\kappa')$ is
fixed-parameter tractable, then so is $(Q,\kappa)$.

\medskip
We also need some notions from parameterized counting complexity. As
they are only required in Section~\ref{sec:count}, we will introduce
them there.

\medskip
\subsection*{Graphs} We only consider \emph{simple} graphs, that is,
finite nonempty undirected graphs without loops and parallel edges.
Every graph $G= (V,E)$ is thus determined by a nonempty vertex set
$V$ and an edge set $E\subseteq \binom{V}{2}$. For an edge
$\{u,v\}\in E$ we say that $u$ is \emph{adjacent} to $v$, and vice
versa. Often we also use $V(G)$ and $E(G)$ to denote the vertex set
and the edge set of $G$, respectively.

\medskip
Let $G=(V,E)$ be a graph. For every vertex $v\in V$ the set $N^G(v)$
contains all vertices in $G$ that are adjacent to $v$, i.e., $N^G(v)
:= \big\{u\bigmid \{u,v\}\in E\big\}$. Moreover, for every
$S\subseteq V$ we let $N^{G}(S) :=\bigcup_{v\in S} N^G(v)$. Note the
degree of $v$, written $\deg^G(v)$, is $|N^G(v)|$. If $\deg^G(v)=0$,
then $v$ is an \emph{isolated} vertex. The distance $d^G(u,v)$
between two vertices $u,v\in V$ is the length of a shortest path
from $u$ to $v$ in the graph $G$.
If it is clear from the context, we omit the superscript $G$ in the
above notations and write $N(v)$, $\deg(v)$, etc., instead.

\medskip
Every nonempty subset $S\subseteq V(G)$ induces a subgraph $G[S]$
with the vertex set $S$ and the edge set $E(G[S]):= \binom{S}{2}\cap
E(G)$. Consequently, a graph $H$ is an \emph{induced subgraph of
$G$} if $H= G[V(H)]$. Recall that $H$ is a \emph{$k$-edge} induced
subgraph of $G$ for $k:= |E(H)|$.

Again, let $S$ be a set of vertices in $G$. Then $S$ is a
\emph{clique}, if for every $u,v \in S$ we have either $u=v$ or
$\{u,v\}\in E(G)$. On the other hand, the set $S$ is an
\emph{independent set} in $G$, if $\{u,v\}\notin E(G)$ for all $u,v
\in S$. For every $k \in \mathbb N$, there exists a constant
$\mathscr R_k$, known as the \emph{Ramsey number}, such that every
graph $G$ with $|V(G)| \ge \mathscr R_k$ has either a clique of size
$k$ or an independent set of size $k$. It is well-known that
$\mathscr R_k < 2^{2\cdot k}$ for every $k\in \mathbb N$.

\subsection*{Relational structures and first-order logic}
A \emph{vocabulary} $\tau$ is a finite set of relation symbols. Each
relation symbol has an \emph{arity}. A \emph{structure}~$\str{A}$ of
vocabulary $\tau$, or simply structure, consists of a nonempty set
$A$ called the \emph{universe}, and an interpretation
$R^{\str{A}}\subseteq A^r$ of each $r$-ary relation symbol $R \in
\tau$. For example, a graph $G$ can be identified with a structure
$\mathcal A(G)$ of vocabulary $\tau_{\rm graph}:= \{E\}$ with the
binary relation symbol $E$ such that $A(G):= V(G)$ and $E^{\mathcal
A(G)}:= \big\{(u,v) \bigmid \{u,v\}\in E(G)\big\}$.

The \emph{disjoint union} of two $\tau$-structures $\mathcal A_1$
and $\mathcal A_2$ is again a $\tau$-structure, denoted by $\mathcal
A_1\, \dot\cup\, \mathcal A_2$, whose universe is $A_1\, \dot\cup\,
A_2$, and where for each relation symbol $R\in \tau$ we let
$R^{\mathcal A_1\, \dot\cup\, \mathcal A_2}:= R^{\mathcal A_1}\,
\dot\cup\, R^{\mathcal A_2}$.

Let $\mathcal A$ be a structure of a vocabulary $\tau$. Then the
\emph{Gaifman graph} of $\mathcal A$ is $G(\mathcal A):=(V,E)$ with
$V:=A$ and
\begin{align*}
E:= \big\{\{a,b\} \bigmid & \text{$a,b\in A$ with $a\ne b$,
 and there exists an $R\in \tau$} \\
 & \quad \text{ and a tuple $(a_1,\ldots, a_r)\in R^{\mathcal A}$ with
 $\{a,b\}\subseteq \{a_1,\ldots, a_r\}$}\big\}.
\end{align*}
Note any unary relation in $\mathcal A$ has no influence on $E$.

Let $r\in \mathbb N$ and $a\in A$. Then the \emph{$r$-neighborhood}
of $a$ is $N^{\mathcal A}_r(a) := \big\{b\in A\bigmid d^{G(\mathcal
A)}(a,b)\le r\big\}$. Moreover, the structure $\mathcal N^{\mathcal
A}_r(a)$ induced by the $r$-neighborhood of $a$ has universe
$N^{\mathcal A}_r(a)$, and for each $r$-ary relation symbol $R\in
\tau$ the interpretation $\big\{(a_1,\ldots, a_r) \in R^{\mathcal A}
\bigmid a_1,\ldots,a_r \in N^{\mathcal A}_r(a)\big\}$.

Formulas of first-order logic of vocabulary $\tau$ are built up from
atomic formulas $x=y$ and $Rx_1 \ldots x_r$ where $x,y,x_1,\ldots,
x_r$ are variables and $R\in\tau$ is of arity $r$, using the boolean
connectives and existential and universal quantification. To give an
example, for every $k\in \mathbb N^+$ let
\[
\textit{is}_k := \exists x_1 \ldots \exists x_k
 \left(\bigwedge_{1\le i<j \le k} (\neg x_i= x_j \wedge \neg Ex_ix_j)\right).
\]
Then a graph $G$ has an independent set of size $k$ if and only if
$\mathcal A(G) \models \textit{is}_k$.

\subsection*{Tree-width and local tree-width}
We assume that the reader is familiar with the notion of
\emph{tree-width} $\tw(G)$ of a graph $G$. Recall that the
tree-width $\tw(\mathcal A)$ of a structure $\mathcal A$ is simply
$\tw(G(\mathcal A))$, that is, the tree-width of the Gaifman graph
of $\mathcal A$. In fact, to understand most parts of our proofs and
algorithms, it is sufficient to know that
\begin{enumerate}
\item[(T)]\label{page:T1} for every structure $\mathcal A$ we
    have $\tw(\mathcal A)< |A|$.
\end{enumerate}


\medskip
Now we are ready to define the \emph{local tree-width} of a
structure $\mathcal A$. For every $r\in \mathbb N$ let
\[
\ltw(\mathcal A, r)
 := \max\left\{\tw\left(\mathcal N^{\mathcal A}_r(a)\right)
 \bigmid a\in A\right\}.
\]
Let $g:\mathbb N\times \mathbb N\to \mathbb N$ be a function and
$p\in \mathbb N$. We say a structure $\mathcal A$ has \emph{local
tree-width bounded by $g$ with respect to $p$} if $\ltw(\mathcal A,
r)\le g(r,p)$ for every $r\in \mathbb N$. This slightly generalizes
the usual notion of local tree-width bounded by a \emph{unary}
function~\cite{frigro}.

\section{Some easy positive instances}\label{sec:comb}

\begin{defn}[\bf independent set matching structure]\label{def:ismatch}
Let $k\in \mathbb N$ and $G=(V,E)$ be a graph. Moreover let $u_1,
\ldots, u_k, v_1, \ldots, v_k$ be $2\cdot k$ vertices in $G$ such
that:
\begin{itemize}
\item[(IM1)] For every $i,j \in [k]$ we have $\{u_i,v_j\} \in E$
    if and only if $i=j$.

\item[(IM2)] $\{u_1,\ldots, u_k\}$ is an independent set in $G$.
\end{itemize}
Then $G$ contains a \emph{$k$-independent-set-matching structure} on
$u_1,\ldots, u_k$, $v_1,\ldots, v_k$.
\end{defn}

\begin{lem}\label{lem:ismatch}
Let $k\in \mathbb N$. Every graph containing a
$k$-independent-set-matching structure has a $k$-edge induced
subgraph.
\end{lem}

\proof The case for $k=0$ is trivially true. So assume $k\ge 1$ and
$G$ contains a $k$-independent-set-matching structure on the
vertices $u_1, \ldots, u_k$, $v_1, \ldots, v_k$.

We choose the maximum $k'\le k$ such that
\[
\ell:= \Big|E\big(G[\{v_1, \ldots, v_{k'}\}]\big)\Big| \le k.
\]
If $k'=k$, then $G[V']$ with $V':=\big\{u_1, \ldots,
u_{k-\ell}\big\} \cup\big\{v_1, \ldots, v_k\big\}$ is a $k$-edge
induced subgraph of $G$.

Otherwise, $k' < k$. In particular, $\Big|E\big(G[\{v_1, \ldots,
v_{k'}, v_{k'+1}\}]\big)\Big|> k$. As $v_{k'+1}$ can contribute at
most $k'$ many new edges, we have $\ell + k' > k$, i.e., $k-\ell <
k'$. Then $G[V']$ with $V':=\big\{u_1, \ldots, u_{k-\ell}\big\}\cup
\big\{v_1, \ldots, v_{k'}\big\}$ is a $k$-edge induced subgraph of
$G$. \proofend

\medskip
\begin{defn}[\bf clique matching structure]\label{def:cliquematch}
Let $k\in \mathbb N$, $G=(V,E)$ be a graph and $u_1, \ldots, u_k,
v_1, \ldots, v_k$ pairwise distinct vertices in $G$ such that:
\begin{itemize}
\item[(CM1)] For every $i,j \in [k]$ we have $\{u_i,v_j\} \in E$
    if and only if $i=j$.

\item[(CM2)] $\{u_1,\ldots, u_k\}$ is a clique in $G$.
\end{itemize}
Then $G$ contains a \emph{$k$-clique-matching structure} on $u_1,
\ldots, u_k$, $v_1, \ldots, v_k$.
\end{defn}

\begin{lem}\label{lem:cliquematching}
Let $k\in \mathbb N$ and $G$ be a graph containing a
$k$-clique-matching structure. Then there is a $k$-edge induced
subgraph in $G$.
\end{lem}

\proof The cases for $k\le 2$ are trivial. So we consider $k\ge 3$.
Let $k_0$ be maximum with $\binom{k_0}{2} \le k$ and set $r:=
k-\binom{k_0}{2}$. It is easy to verify that $k\ge k_0+r$ by $k\ge
3$ and $k_0> r$. Now assume $G$ contains a
$k$-clique-matching-structure on the vertices $u_1, \ldots, u_k$,
$v_1,\ldots, v_k$. Then, we choose the maximum $r'\le r$ such that
\[
\ell:= \Big|E\big(G[\{v_1, \ldots, v_{r'}\}]\big)\Big| \le r.
\]
If $r'=r$, then $G[V']$ with $V':= \big\{v_1, \ldots, v_{r}\big\}
\cup \big\{u_1, \ldots,
u_{r-\ell},u_{r+1},\ldots,u_{k_0+\ell}\big\}$ is a $k$-edge induced
subgraph of $G$.
Otherwise, $r' < r$ and by the maximality of $r'$ we have
$\Big|E\big(G[\{v_1, \ldots, v_{r'}, v_{r'+1}\}]\big)\Big|> r$. As
$v_{r'+1}$ can add at most $r'$ many new edges, we have $\ell+ r'>
r$, or equivalently $r-\ell < r'$. It follows that $G[V']$ with
$V':= \big\{v_1, \ldots, v_{r'}\big\} \cup \big\{u_1, \ldots,
u_{r-\ell},u_{r'+1},\ldots,u_{r'+k_0-r+\ell}\big\}$ has exactly $k$
edges. \proofend

\medskip
\begin{defn}[\bf apex structure]\label{def:apex}
Let $k\in \mathbb N$, $G= (V,E)$ be a graph, $A,B\subseteq V$, and a
vertex $v_0 \in V$ which satisfy the following conditions:
\begin{itemize}
\item[(A1)] $A, B$ are disjoint with $|A|\ge k$ and $|B|\ge
    \mathscr R_k$.

\item[(A2)] $A$ is a clique in $G$.

\item[(A3)] $\{u,v_0\}\in E$ for every $u \in A$ and
    $\{v,v_0\}\notin E$ for every $v\in B$. \big(Note this
    implies that $v_0\notin A$ but possibly $v_0\in B$.\big)

\item[(A4)] $\{u,v\}\in E$ for every $u\in A$ and $v\in B$.
\end{itemize}
Then we say that $G$ contains a \emph{$k$-apex structure} on $v_0$,
$A$ and $B$.
\end{defn}

\begin{lem}\label{lem:apex}
Let $k\in \mathbb N$ and $G$ be a graph. If $G$ contains a $k$-apex
structure, then it has a $k$-edge induced subgraph.
\end{lem}

\proof The case for $k\le 1$ is trivially true. So let $k\ge 2$.
Moreover, let $v_0, A, B$ be as stated in Definition~\ref{def:apex}.
Since $|B| \ge \mathscr R_k$, $G[B]$ contains either a clique of
size $k$ or an independent set of size $k$.

If $G[B]$ contains an independent set $B' \subseteq B$ with
$|B'|=k$. Then for every $u\in A$ the induced subgraph $G\big[B'
\cup \{u\}\big]$ has exactly $k$ edges by (A4).

Now assume that there is a clique $B'$ in $G[B]$ of size $k$.
Observe by (A3) and $k\ge 2$, we have $v_0\notin (A\cup B')$.
Furthermore, it is easy to see that we can write $k = \binom{k_0}{2}
+ r$ for some appropriate $k\ge k_0\ge r$.

\medskip
We select arbitrary subsets $A' \subseteq A$ and $B'' \subseteq B'$
with $|A'|= r$ and $|B''|= k_0-r$. Then it is straightforward to
check that $G\big[A' \cup B'' \cup \{v_0\}\big]$ has exactly $k$
edges. \proofend

\medskip
\begin{lem}[\bf three cliques]\label{lem:3cliques}
Let $k \in \mathbb N$ and $G= (V,E)$ be a graph. Assume there exists
three subsets $S_1, S_2, S_3$ such that:
\begin{itemize}
\item $S_1, S_2, S_3$ are three disjoint cliques in $G$, all of
    size $k$.

\item There are no edges between any distinct $S_i$ and $S_j$.
\end{itemize}
Then $G$ has a $k$-edge induced subgraph.
\end{lem}

It is easy to see that Lemma~\ref{lem:3cliques} is a direct
consequence of Gauss' famous Eureka Theorem~\cite{and}.

\begin{theo}
For every $k \in \mathbb N$ there exist $k_0,k_1,k_2 \in \mathbb N$
such that
\[
k = \binom{k_0}{2}+ \binom{k_1}{2}+ \binom{k_2}{2}.
\]
\end{theo}

\medskip

\begin{lem}[\bf large independent set]\label{lem:is}
Let $k\in \mathbb N^+$ and $G= (V,E)$ be a graph without isolated
vertices. If $G$ contains an independent set of size $(k-1)^2+ 1$,
then it has a $k$-edge induced subgraph.
\end{lem}

To prove the above lemma, we need some further preparation.

\begin{lem}\label{lem:bi1}
Let $m,n \in \mathbb N^+$ and $G= (V,E)$ be a graph. Furthermore,
let $A, B\subseteq V$ be disjoint such that $\big|N(u) \cap B\big|
\ge 1$ for every $u\in A$. If $|A|> (m- 1)(n- 1)$, then
\begin{itemize}
\item[(i)] either there are $m$ vertices $u_1, \ldots, u_m$ in
    $A$ and a vertex $v$ in $B$ with $\{u_i,v\}\in E$ for every
    $i\in [m]$, 

\item[(ii)] or there are $n$ vertices $u_1, \ldots, u_{n}$ in
    $A$ and $n$ vertices $v_1,\ldots, v_{n}$ in $B$ such that
    for all $i,j\in [n]$ we have $\{u_i, v_j\}\in E$ if and only
    if $i=j$.
\end{itemize}
\end{lem}

\proof Let $s:= |B|$. We prove by induction on $s$ and $n$. If $n=
1$, then (ii) is trivially true. And if $s= 1$ and $n> 1$, then
clearly (i) holds.

Now assume both $s> 1$ and $n> 1$. If there exists a vertex $v\in B$
with $\big|N(v) \cap A\big| \ge m$, then we can easily achieve (i).
So assume now that
\begin{equation}\label{eq:NvA}
\text{for every $v\in B$ we have $\big|N(v) \cap A\big| \le m-1$}.
\end{equation}
Choose an arbitrary vertex $v\in B$ and let $B':= B\setminus \{v\}$.
If for every $u\in A$ we have $\big|N(u) \cap B'\big|\ge 1$, then
the result follows from the induction hypothesis on $A$ and $B'$
with $|B'|= s-1$. Otherwise, there exists a vertex $u\in A$ such
that $N(u)\cap B'= \emptyset$, i.e., $N(u)\cap B = \{v\}$. Let $A':=
A\setminus N(v)$. By~\eqref{eq:NvA} it holds that $|A'|> (m- 1)(n-
2)$. Then by induction hypothesis on
\[
 A\gets A', B\gets B', m\gets m, \ \text{and}\ n\gets n-1,
\]
together with~\eqref{eq:NvA}, the property (ii) holds for $A'$,
$B'$, and $n-1$. That is, there are $n-1$ vertices $u_1, \ldots,
u_{n-1}$ in $A'$ and $n-1$ vertices $v_1,\ldots, v_{n-1}$ in $B'$
such that for all $i,j\in [n-1]$ we have $\{u_i, v_j\}\in E$ if and
only if $i=j$.
As $N(u)\cap B' = N(v)\cap A'= \emptyset$, by taking $u_{n}:= u$ and
$v_{n}:= v$, we have $\{u_i, v_j\}\in E$ if and only if $i=j$, for
every $i,j \in [n]$. \proofend

%

\medskip
\noindent \textit{Proof of Lemma~\ref{lem:is}:} Let $S \subseteq V$
be an independent set in $G$ with $|S|> (k-1)^2$. Since $G$ has no
isolated vertex, $|N(u) \cap N(S)| \ge 1$ for every $u\in S$. So we
can apply Lemma~\ref{lem:bi1} on
\[
A\gets S, B\gets N(S), m\gets k,\ \ \text{and}\ m\gets k.
\]
If (i) holds, then we have an induced $k$-star of exactly $k$ edges.
Otherwise, we have (ii). Hence, there exist vertices $u_1, \ldots,
u_k\in S$ and $v_1, \ldots, v_k \in N(S)$ such that $G$ contains
$k$-independent-set-matching structure on those vertices. The result
follows from Lemma~\ref{lem:ismatch}. \proofend

\medskip

\begin{defn}\label{def:V1d}
Let $G=(V,E)$ be a graph and $d\in \mathbb N$. We define
\begin{equation*}
V^G_{[1,d]} := \big\{v\in V\mid 1\le \deg(v)\le d\big\}.
\end{equation*}
\end{defn}

\begin{lem}[\bf sufficiently many small degree
vertices]\label{lem:smalldegree} Let $d,k \in \mathbb N^+$ and $G=
(V,E)$ be a graph. If $\left|V^G_{[1,d]}\right|> (d+1)\cdot
(k-1)^2$, then $G$ contains a $k$-edge induced subgraph.
\end{lem}

\proof Let $G'=(V',E')$ be the graph resulting by removing all
isolated vertices from $G$. Then, by Lemma~\ref{lem:is} it suffices
to show that $G'$ contains an independent set $S$ of size
$(k-1)^2+1$. In fact, such a set $S$ can be constructed by
repeatedly picking vertices from $V_{[1,d]}\subseteq V'$ and
removing their neighbors. \proofend

\medskip

\begin{rem}
An immediate consequence of Lemma~\ref{lem:smalldegree} is that
\pEIS\ is solvable in time $2^{O(d\cdot k^2)}+ n^{2}$ on graphs of
degree $\le d$.
\end{rem}

\subsection{A further combinatorial
lemma}\label{subsec:furtherlemma} For later purpose, we need a
generalization of Lemma~\ref{lem:bi1}.

\begin{lem}\label{lem:bi}
Let $m,n,p\in \mathbb N^+$ and $G= (V,E)$ be a graph. Furthermore,
let $A, B\subseteq V$ be disjoint such that $\big|N(u) \cap B\big|
\ge p$\, for every $u\in A$. If $|A| > (m-1)(n-1)^p$, then
\begin{itemize}
\item[(i)] either there are $m$ vertices $u_1, \ldots, u_m$ in
    $A$ and $p$ vertices $v_1, \ldots, v_p$ in $B$ with
    $\{u_i,v_j\}\in E$ for every $i\in [m]$ and $j\in [p]$,

\item[(ii)] or there are $n$ vertices $u_1, \ldots, u_n$ in $A$
    and $n$ vertices $v_1,\ldots, v_n$ in $B$
    such that for all $i,j\in [n]$ we have $\{u_i, v_j\}\in E$
    if and only if $i=j$.
\end{itemize}
\end{lem}

\proof We proceed by induction on $p$. The case $p=1$ is precisely
Lemma~\ref{lem:bi1}. So let $p> 1$. We apply Lemma~\ref{lem:bi1}
on
\[
m\gets (m-1)(n-1)^{p-1}+1\quad\text{and}\quad n\gets n .
\]
Thus
\begin{itemize}
\item[(a)] either there are $(m-1)(n-1)^{p-1}+1$ vertices $u_1,
    \ldots, u_{(m-1)(n-1)^{p-1}+1}$ in $A$ and a vertex $v$ in
    $B$ with $\{u_i,v\}\in E$ for every $i\in
    \big[(m-1)(n-1)^{p-1}+1\big]$,

\item[(b)] or there are $n$ vertices $u_1, \ldots, u_n$ in $A$
    and $n$ vertices $v_1,\ldots, v_n$ in $B$ such that for all
    $i,j\in [n]$ we have $\{u_i, v_j\}\in E$ if and only if
    $i=j$.
\end{itemize}
Clearly (b) is exactly (ii). So we assume that (a) holds. Let
\[
A' := \big\{u_1, \ldots, u_{(m-1)(n-1)^{p-1}+1}\big\}, \quad
B' := B \setminus \{v\}, \quad
m' := m, \quad
n' := n, \quad \text{and}\quad
p' := p-1.
\]
It is easy to verify that we can apply the induction hypothesis on
\[
A \gets A', B \gets B', m\gets m', n\gets n',\ \text{and}\ p\gets p'.
\]
If (ii) holds for $A'$, $B'$, and $n'$, then it holds for $A$, $B$,
$n$, too. Otherwise there are $m$ vertices $u'_1, \ldots, u'_m$ in
$A'\subseteq A$ and $p- 1$ vertices $v'_1, \ldots, v'_{p-1}$ in
$B'\subseteq B$ with $\{u'_i,v'_j\}\in E$ for every $i\in [m]$ and
$j\in [p-1]$.

Recall now (a) is true for the vertices in $A$ and the vertex $v$ in
$B$. Therefore, $\{u'_i, v\}\in E$ for every $i\in [m]$. Then (i)
holds for $u'_1, \ldots, u'_m\in A$, $v'_1, \ldots, v'_{p-1},v\in
B$, $m$, and $p$ by $v\in B \setminus B'$. \proofend

\section{Easy instances by model-checking}\label{sec:logic}
In this section we show the fixed-parameter tractability of \pEIS\
on some restricted classes of graphs via the model-checking problem
for first-order logic.

\medskip
As mentioned in the Introduction, the following is a generalization
of a well-known result due to Frick and Grohe~\cite{frigro}.
\begin{theo}\label{theo:mcltwp0}
For every computable function $g: \mathbb N\times \mathbb N \to
\mathbb N$ the problem
\npprob{$\pMCLTWFO g$}{A structure $\mathcal A$, $p\in \mathbb N$
and an FO-sentence $\varphi$ such that $\mathcal A$ has local
tree-width bounded by $g$ with respect to $p$}{$p+|\varphi|$}{Decide
whether $\mathcal A\models \varphi$.}
is fixed-parameter tractable.
\end{theo}
For the sake of completeness we include a proof in the appendix.

\medskip
\begin{defn}[\bf degree-extreme graph]\label{def:degreeextreme}
Let $d\in \mathbb N$ and $G=(V,E)$ be a graph. If $\deg(v) \le d$ or
$\deg(v)\ge |V|- 1 - d$ for every $v\in V$, then the graph $G$ is
\emph{$d$-degree-extreme}.
%
For example, let $n\in \mathbb N$, then an $n$-star is
$d$-degree-extreme for every $d\ge 1$.
\end{defn}

\medskip
Now we translate every degree-extreme graph to a finite structure
over the vocabulary $\tau_{\rm des} := \{P, R\}$ where $P$ is a
unary relation symbol and $R$ a binary relation symbol.

\begin{defn}[\bf degree-extreme structure]\label{def:des}
Let $d\in \mathbb N$ and $G= (V,E)$ be a $d$-degree-extreme graph.
We set $V^G_{\le d}:= \big\{v\in V\bigmid \deg(v) \le d\big\}$. Then
$\mathcal A:= \mathcal A(G,d)$ is a $\tau_{\rm des}$-structure
defined by $A:= V$, $P^{\mathcal A}:= V^G_{\le d}$, and
\begin{align*}
R^{\mathcal A}:=
 \Big\{(u,v) \Bigmid & \ \text{$\{u,v\}\in E$
  and \big($u\in V^G_{\le d}$ or $v\in V^G_{\le d}$\big)}\Big\} \\
 & \cup \Big\{(u,v) \Bigmid \text{$\{u,v\} \notin E$,
  $u,v \in V\setminus V^G_{\le d}$ and $u\ne v$} \Big\}.
\end{align*}
%
Basically, $\mathcal A(G,d)$ has the same vertex set as $G$, keeps
the edges between two small degree vertices and the edges between a
small degree vertex and a large degree one, and takes the complement
of remaining edges between large degree vertices.
%
\end{defn}

\medskip
\begin{lem}\label{lem:des}
There is a computable function $h_0:\mathbb N\times \mathbb N\times
\mathbb N^+\to \mathbb N^+$ such that for every $d\in \mathbb N$,
$k\in \mathbb N^+$ and every $d$-degree-extreme graph $G$ we have
\begin{itemize}
\item[(i)] either $\left|V^G_{[1,d]}\right| > (d+1)\cdot
    (k-1)^2$, \big(hence, by Lemma~\ref{lem:smalldegree}, $G$
    has a $k$-edge induced subgraph\big),

\item[(ii)] or for the structure $\mathcal A:= \mathcal A(G,d)$
    as defined in Definition~\ref{def:des} we have
    $\ltw\left(\mathcal A, r\right)\le h_0(r,d,k)$ for every
    $r\in \mathbb N$.
\end{itemize}
\end{lem}

\proof We assume that (i) is not true, i.e.,
$\left|V^G_{[1,d]}\right| \le (d+1)\cdot (k-1)^2$. For every $v\in A
= V(G)$
it is easy to verify that $\deg^{G(\mathcal A)}(v) \le d+ (d+1)\cdot
(k-1)^2$. Together with (T)\big(see page~\pageref{page:T1}\big) we
conclude
\begin{align*}
\tw\left(\mathcal N^{\mathcal A}_r(v)\right)
 & < \big|N^{\mathcal A}_r(v)\big|
   \le \sum_{i=0}^{r} \left(d+ (d+1)\cdot (k-1)^2\right)^i.
\end{align*}
Thus we can define the desired function $h_0$ accordingly. \proofend


\medskip
\begin{defn}\label{def:desset}
Recall the vocabulary of degree-extreme structures is $\tau_{\rm
des}=\{P,R\}$. We let
\begin{align*}
\textit{edge}(x,y):=  & \big(Rxy \wedge (Px \vee Py)\big)
 \vee (\neg Rxy \wedge \neg Px \wedge \neg Py).
\end{align*}
Moreover, let $H= (V,E)$ be a graph. We assume that $V=[\ell]$ for
some $\ell\in \mathbb N$. We define
\begin{align*}
\textit{induced}_H:= \exists x_1\ldots \exists x_{\ell}
 \left(\bigwedge_{1\le i<j \le \ell} \neg x_i= x_j \wedge
  \bigwedge_{\{i,j\}\in E} \textit{edge}(x_i,x_j) \wedge
  \bigwedge_{\{i,j\}\in \binom{V}{2}\setminus E} \neg \textit{edge}(x_i,x_j)\right).
\end{align*}
\end{defn}

\medskip
Then the following lemma is straightforward.
\begin{lem}\label{lem:desmc1}
Let $d\in \mathbb N$ and $G$ be a $d$-degree-extreme-graph. For
every graph $H$ we have
\begin{eqnarray*}
\text{$G$ contains an induced subgraph isomorphic to $H$}
 & \iff & \text{$\mathcal A(G,d)
\models \textit{induced}_{H}$}.
\end{eqnarray*}
\end{lem}

\begin{prop}\label{prop:mcdes}
Let $D: \mathbb N\to \mathbb N$ be a computable function. Then the
problem
\pprob[9]{A graph $G$ and $k\in \mathbb N$ such that $G$ is
$D(k)$-degree-extreme}{$k$}{Decide whether $G$ contains a $k$-edge
induced subgraph.}
is fixed-parameter tractable.
\end{prop}

\proof We only consider $k\in \mathbb N^+$ and let $G=(V,E)$ be a
$D(k)$-degree-extreme graph. Moreover, let $\mathcal A:= \mathcal
A(G,D(k))$. By Lemma~\ref{lem:des} we can assume that
\[
\ltw(\mathcal A, r) \le h_0(r, D(k), k).
\]
That is, the structure $\mathcal A$ has local tree-width bounded by
the function $g(r,k):= h_0(r, D(k), k)$ with respect to $k$.

Then we define the following \FO-sentence
\[
\textit{induced}_k:= \bigvee_{\substack{\text{$H$ has no isolated vertex}\\[1mm]
  \text{and $|E(H)|= k$}}}
  \textit{induced}_H.
\]
It follows that $G$ has an induced subgraph of exactly $k$ edges if
and only if $\mathcal A \models \textit{induced}_k$. Note the
structure $\mathcal A$ can be computed in \fpt\ time, and the
sentence $\textit{induced}_k$ can be computed from $k$. Hence,
$(G,k) \mapsto \big(\mathcal A, k, \textit{induced}_k\big)$ gives an
\fpt-reduction to $\pMCLTWFO g$. The result then follows from
Theorem~\ref{theo:mcltwp0}. \proofend

\begin{rem}\label{rem:mcdestime}
A careful analysis of the above algorithm shows that its running
time in terms of the parameter $k$ is at least of the order of
$2^{\Theta(D(k))}$.
\end{rem}

\medskip

\begin{defn}[\bf bridge]\label{def:bridge}
Let $d,b\in \mathbb N$. Moreover let $G = (V,E)$ be a graph such
that:
\begin{itemize}
\item[(B1)] $V= V_1\cup V_2$ for some disjoint $V_1$ and $V_2$.

\item[(B2)] $G[V_1]$ and $G[V_2]$ are both $d$-degree-extreme.

\item[(B3)] There exists a subset $B\subseteq V$ with $|B| = b$
    such that for every edge $\{u,v\}$ with $u \in V_1$ and
    $v\in V_2$ we have either $u\in B$ or $v\in B$.
\end{itemize}
Then $(G,V_1, V_2, B)$ is a \emph{$(d,b)$-bridge} (of the two
degree-extreme graphs).
\end{defn}

\medskip
Similarly to degree-extreme graphs, we translate every bridge to a
finite structure. To that end, for every $b\in \mathbb N$ let
\[
\tau_{{\rm bridge},b} := \big\{U_1, U_2, P, R,
 F_1, \ldots, F_b, C_1, \ldots, C_b\big\},
\]
where all symbols are unary except the binary $R$.

\medskip

\begin{defn}[\bf bridge structure]\label{def:brigestructure}
Let $d,b\in \mathbb N$, $G= (V,E)$ be a graph and $V_1, V_2,
B\subseteq V$ with $B= \big\{v_1, \ldots, v_b\big\}$ such that
$(G,V_1,V_2,B)$ is a $(d,b)$-bridge of two $d$-degree-extreme graphs
$G[V_1]$ and $G[V_2]$. Then we define the corresponding $\tau_{{\rm
bridge},b}$-structure
\begin{equation}\label{eq:D}
\mathcal D:= \mathcal D(G,V_1,V_2,B,d)
 := \Big(\mathcal A(G[V_1],d)\; \dot\cup\; \mathcal A(G[V_2],d),
 U^{\mathcal D}_1, U^{\mathcal D}_2,
 F^{\mathcal D}_1, \ldots, F^{\mathcal D}_b,
 C^{\mathcal D}_1, \ldots, C^{\mathcal D}_b\Big),
\end{equation}
where $U^{\mathcal D}_1:= V_1$, $U^{\mathcal D}_2:= V_2$ and for
every $i\in [b]$
\[
F^{\mathcal D}_i:= \{v_i\}, \ C^{\mathcal D}_i:= \big\{u\in V\bigmid
\{u,v_i\}\in E\big\}.
\]
%
That is, the bridge structure consists of two degree-extreme
structures, plus all the edges between them encoded by $2\cdot b$
unary relations.
%
\end{defn}

\medskip
\begin{lem}\label{lem:bridge}
Let $d\in \mathbb N$, $k\in \mathbb N^+$, $G=(V,E)$ be a graph and
$V_1,V_2,B\subseteq V$ such that $(G,V_1,V_2,B)$ is a
$(d,|B|)$-bridge. Moreover, let $\mathcal D:= \mathcal
D(G,V_1,V_2,B,d)$. Then one of the following conditions is
satisfied.
\begin{itemize}
\item[(i)] $\left|V^{G[V_1]}_{[1,d]}\right| > (d+1)\cdot
    (k-1)^2$.

\item[(ii)] $\left|V^{G[V_2]}_{[1,d]}\right| > (d+1)\cdot
    (k-1)^2$.

\item[(iii)]  $\ltw(\mathcal D,r) \le h_0(r,d,k)$ for every
    $r\in \mathbb N$, where the function $h_0$ is defined in
    Lemma~\ref{lem:des}.
\end{itemize}
Observe in cases (i) and (ii), by Lemma~\ref{lem:smalldegree},
$G[V_1]$ or $G[V_2]$ and hence $G$ has a $k$-edge induced subgraph.
\end{lem}

\proof Assume that neither (i) nor (ii) holds. Let $v\in D= V$,
$r\in \mathbb N$ and consider the structure $\mathcal N^{\mathcal
D}_r(v)$. Observe that all unary relations $U^{\mathcal D}_1,\ldots,
C^{\mathcal D}_b$ have no impact on the tree-width of $\mathcal
N^{\mathcal D}_r(v)$, i.e.,
\[
\tw\left(\mathcal N^{\mathcal D}_r(v)\right)
 = \tw\Big(\mathcal N^{\mathcal A(G[V_1],d)\,
\dot\cup\, \mathcal A(G[V_2],d)}_r(v)\Big)
\]
by~\eqref{eq:D}. Hence
\[
\tw\left(\mathcal N^{\mathcal D}_r(v)\right) =
\begin{cases}
\tw\left(\mathcal N^{\mathcal A(G[V_1],d)}_r(v)\right),
 & \text{if $v\in V_1$} \\
\tw\left(\mathcal N^{\mathcal A(G[V_2],d)}_r(v)\right),
 & \text{if $v\in V_2$}.
\end{cases}
\]
Then (iii) follows from Lemma~\ref{lem:des}. \proofend

\medskip
\begin{defn}\label{def:bridgeset}
For every $b\in \mathbb N$ let
\begin{align*}
\textit{edge}^2_b(x,y):=
 & \quad \big(U_1x\wedge U_1y \wedge \textit{edge}(x,y)\big)
  \vee \big(U_2x\wedge U_2y \wedge \textit{edge}(x,y)\big) \\
 &\qquad  \vee \bigvee_{i\in [b]} \big((F_ix \wedge C_iy)
       \vee (F_iy \wedge C_ix)\big).
\end{align*}
Recall the formula $\textit{edge}(x,y)$ is defined in
Definition~\ref{def:desset}.

\medskip
Then for every graph $H= (V,E)$, where $V=[\ell]$ for some $\ell\in
\mathbb N$, we define
\begin{align*}
\textit{induced}^2_{b,H}:= \exists x_1\ldots \exists x_{\ell}
 \left( \bigwedge_{1\le i<j \le \ell} \neg x_i= x_j \wedge
  \bigwedge_{\{i,j\}\in E} \textit{edge}^2_b(x_i,x_j) \wedge
  \bigwedge_{\{i,j\}\in \binom{V}{2}\setminus E} \neg \textit{edge}^2_b(x_i,x_j)\right).
\end{align*}
\end{defn}

\medskip
\begin{lem}\label{lem:bsmc}
Let $d,b\in \mathbb N$, $G= (V,E)$ a graph and $V_1,V_2, B\subseteq
V$ such that $(G,V_1,V_2,B)$ is a $(d,b)$-bridge. Then for every
graph $H$ we have
\begin{eqnarray*}
\text{$G$ contains an induced subgraph isomorphic to $H$}
 & \iff & \text{$\mathcal D(G,V_1,V_2,B,d)
\models \textit{induced}^2_{b,H}$}.
\end{eqnarray*}
\end{lem}
We omit the trivial proof.

\medskip

\begin{prop}\label{prop:mcbs}
Let $D: \mathbb N\to \mathbb N$ be a computable function. Then the
problem
\pprob{A graph $G=(V,E)$, $V_1,V_2,B\subseteq V$ and $k\in \mathbb
N$ such that $(G,V_1,V_2,B)$ is a
$(D(k),|B|)$-bridge}{$k+|B|$}{Decide whether $G$ contains a $k$-edge
induced subgraph.}
is fixed-parameter tractable.
\end{prop}

\proof This is similar to Proposition~\ref{prop:mcdes}.
\proofend

\section{The algorithm}\label{sec:algorithm}
The main component of our \fpt-algorithm for \pEIS\ is the following
procedure that either already solves the problem or decomposes the
given graph into potentially a bridge of two large degree-extreme
graphs (cf. Definition~\ref{def:bridge}).

\bigskip
For every $k\in \mathbb N$ we let
\begin{eqnarray*}
p_k:= 2^{2\cdot k} (> \mathscr R_k).
\end{eqnarray*}

\begin{lem}\label{lem:main}
For every computable function $D:\mathbb N\to \mathbb N$ there is an
\fpt-algorithm $\mathbb A_{D}$ such that for every graph $G=(V,E)$
and every $k\in \mathbb N$ exactly one of following conditions is
satisfied.
\begin{itemize}
\item[(S1)] $G$ is $D(k)$-degree-extreme and $\mathbb A_{D}$
    correctly decides whether $G$ contains a $k$-edge induced
    subgraph.

\item[(S2)] $G$ is not $D(k)$-degree-extreme and $\mathbb A_{D}$
    correctly outputs that $G$ contains a $k$-edge induced
    subgraph.

\item[(S3)] $G$ is not $D(k)$-degree-extreme and $\mathbb A_{D}$
    outputs three subsets $V_1, V_2, B\subseteq V$ such that
    \begin{enumerate}
    \item[(S3.1)] $V= V_1\, \dot\cup\, V_2$ with $|V_1|>
        D(k)$ and $|V_2|> D(k)+ 1$;

    \item[(S3.2)] every edge between $V_1$ and $V_2$ in $G$
        has one vertex in $B$ and $|B| \le
        (p_k-1)^{p_k+1}+ (p_k-1)^2$.

    \end{enumerate}
\end{itemize}
\end{lem}

\proof Let $G=(V,E)$ be a graph and $k\in \mathbb N$. If $G$ is
$D(k)$-degree-extreme, then we apply Proposition~\ref{prop:mcdes} to
achieve (S1). Otherwise let $v_0\in V$ be a vertex with
\begin{equation}\label{eq:v0}
D(k)< \deg(v_0) < |V|-1-D(k).
\end{equation}
Then we set $V_1:= N(v_0)$ and $V_2:=V\setminus V_1$.
By~\eqref{eq:v0} it holds that $|V_1|> D(k)$ and $|V_2|= |V|-|V_1| =
|V|- \deg(v_0) > D(k)+1$, i.e., (S3.1).
%
%
Let
\begin{eqnarray*}
W_1 := \Big\{u \in V_1\Bigmid \big|N(u) \cap V_2\big| \ge
p_k\Big\} & \text{and} &
W_2 := V_1\setminus W_1.
\end{eqnarray*}
Figure~\ref{fig1} illustrates our construction.

\definecolor{LightYellow}{rgb}{1.,1.,0.88}
\definecolor{Peach}      {rgb}{1,.859,.780}
\definecolor{MyGray}     {gray}{0.8}

\newcommand{\curveone}{%
\pscurve[showpoints=false,linewidth=.05]%
  (2.5,12)(2,10)(2,8)(4,4)(7,4)(9,10)
 \pscustom[linewidth=.05,fillstyle=solid,fillcolor=MyGray]{%
 \pscurve(9,10)(8.5,12)(5,15)(2.5,12)
 \pscurve(2.5,12)(4,11.1)(5,10.9)(6,11)(7,11)(9,10)
 }
}
\newcommand{\curvetwo}{%
\pscurve[showpoints=false,linewidth=.05]%
  (19,6)(19,13)(14,14)(13,13)(11,7)
\pscustom[linewidth=.05,fillstyle=solid,fillcolor=MyGray]{%
\pscurve(19,6)(17,7)(15,8)(13,8)(11,7)
\pscurve(11,7)(11.5,3.5)(16.5,3)(19,6) } }

\newcommand{\edges}{%

\psline[linewidth=.1,linearc=.25,linecolor=MyGray]{-}(7,12)(16,12)
\psline[linewidth=.1,linearc=.25,linecolor=MyGray]{-}(3.5,8.5)(16,12)
\psline[linewidth=.1,linearc=.25,linecolor=MyGray]{-}(5.5,5.5)(16,12)
\psdot[dotsize=3pt](16,12) \psdot[dotsize=3pt](3.5,8.5)

\psdot[dotsize=3pt](5.5,5.5)
\psline[linewidth=.1,linearc=.25]{-}(5.5,5.5)(12.5,6.5)
\psdot[dotsize=3pt](12.5,6.5)
\psline[linewidth=.1,linearc=.25]{-}(5.5,5.5)(16,4.5)
\psdot[dotsize=3pt](16,4.5)

\psdot[dotsize=3pt](7,12)
\psline[linewidth=.1,linearc=.25]{-}(7,12)(13,7.2)
\psdot[dotsize=3pt](13,7.2)
\psline[linewidth=.1,linearc=.25]{-}(7,12)(14,9)
\psdot[dotsize=3pt](14,9)
\psline[linewidth=.1,linearc=.25]{-}(7,12)(13,11)
\psdot[dotsize=3pt](13,11)
\psline[linewidth=.1,linearc=.25]{-}(7,12)(15,13.5)
\psdot[dotsize=3pt](15,13.5)

 }

\newcommand{\labels}{%
 \rput(16,15){$V_2$}
 \rput(15,5.5){$N(W_2)\cap V_2$}
 \rput(6,16){$V_1$}
 \rput(5,13){$W_1$}
 \rput(6,8){$W_2$}
 \rput(11.5,12){$\ge q_k$}
 \rput(10,4){$<q_k$}

 \rput(16,12.7){$v_0$}
 }

\psset{unit=0.8em,linewidth=.5em,dimen=middle}
\begin{center}
\begin{figure}
  \centering
  {\small
  \AltClipMode
  \begin{pspicture}*(0.5,0.5)(20,20)
    \curveone
    \curvetwo
    \edges
    \labels
  \end{pspicture}}
  \caption{}\label{fig1}
\end{figure}
\end{center}

\noindent \textit{Claim 1.} If $|W_1|> (p_k-1)^{p_k+1}$, then
$G$ contains a $k$-edge induced subgraph.

\medskip
\noindent \textit{Proof of the claim.} We apply Lemma~\ref{lem:bi}
on
\[
A \gets W_1, B \gets V_2, m\gets p_k, n \gets p_k,\ \text{and}\ p \gets p_k.
\]
So there are $p_k$ vertices $u_1, \ldots, u_{q_k}$ in $W_1$ and
$p_k$ vertices $v_1, \ldots, v_{p_k}$ in $V_2$ such that
\begin{itemize}
\item[(i)] either  $\{u_i,v_j\}\in E$ for every $i,j\in [p_k]$,

\item[(ii)] or for all $i,j\in [p_k]$ we have $\{u_i, v_j\}\in
    E$ if and only if $i=j$.
\end{itemize}

Recall $p_k> \mathscr R_k$, so there is a subset $S\subseteq
\{u_1,\ldots,u_{p_k}\}$ such that $S$ is either an independent set
or a clique. If $S$ is an independent set, then $G\big[S\cup
\{v_0\}\big]$ has exactly $k$ edges. So suppose $S$ is a clique.

Assume that (i) is true, then $G$ contains a $k$-apex structure on
$v_0, S, \{v_1, \ldots, v_{p_k}\}$. Hence, Lemma~\ref{lem:apex}
implies the claim. Otherwise (ii) holds. And say $S= \{u_{i_1},
\ldots, u_{i_k}\}$. Then the graph $G$ contains an
$k$-clique-matching structure on $u_{i_1}, \ldots, u_{i_k}, v_1,
\ldots, v_{k}$. The result follows from
Lemma~\ref{lem:cliquematching}. \hfill$\dashv$

\bigskip \noindent \textit {Claim 2.} If $\big|N(W_2)\cap V_2\big|>
(p_k-1)^2$, then $G$ contains a $k$-edge induced subgraph.

\medskip
\noindent \textit{Proof of the claim.} It is easy to verify that we
can apply Lemma~\ref{lem:bi1} on
\begin{align*}
A\gets N(W_2)\cap V_2, B\gets W_2, m\gets p_k, \ \text{and}\ n\gets p_k.
\end{align*}
So,
\begin{itemize}
\item[(i)] either there are $p_k$ vertices $u_1,\ldots, u_{q_k}$
    in $N(W_2)\cap V_2$ and a vertex $v$ in $W_2$ such that
    $\{u_i, v\}\in E$ for every $i\in [p_k]$,

\item[(ii)] or there are $p_k$ vertices $u_1,\ldots, u_{p_k}$ in
    $N(W_2)\cap V_2$ and $p_k$ vertices $v_1,\ldots, v_{p_k}$ in
    $W_2$ such that for all $i,j\in [p_k]$ we have
    $\{u_i,v_j\}\in E$ if and only if $i=j$.
\end{itemize}
But (i) contradicts our definition of $W_2$, i.e., for every $u\in
W_2$ we have $\big|N(u)\cap V_2\big|<p_k$, therefore (ii) must hold.
Recall $p_k> \mathscr R_{k}$, hence $G\big[\{v_1,\ldots,
v_{p_k}\}\big]$ contains either a clique of size of $k$ or an
independent set of size $k$. Without loss of generality, let
$\{v_1,\ldots, v_k\}\subseteq W_2\subseteq V_1$ be a clique or an
independent set.

For the independent set case, as $v_0\notin V_1$, then
$G\big[\{v_0,v_1,\ldots,v_k\}\big]$ is a $k$-induced subgraph. For
the clique case, $G$ contains a $k$-clique-matching structure on
$u_1,\ldots,u_{k}, v_1,\ldots, v_{k}$. We are done by
Lemma~\ref{lem:cliquematching}. \hfill$\dashv$

\medskip
Let
\begin{equation*}
B := W_1 \cup \big(N(W_2)\cap V_2\big),
\end{equation*}
i.e., the grey area in Figure~\ref{fig1}.  If $|B|> (p_k-1)^{p_k+1}
+ (p_k-1)^2$, then, by Claim~1 and Claim~2, the graph $G$ contains a
$k$-edge induced subgraph, and (S2) follows. Otherwise
\begin{equation*}
|B| \le (p_k-1)^{p_k+1}+(p_k-1)^2.
\end{equation*}
Observe that every edge between $V_1$ and $V_2$ has at least one
vertex in $B$. Thus, we achieve (S3) by outputting $(V_1, V_2, B)$.
\proofend

Finally we are ready to present our \fpt-algorithm for $\pEIS$.

\begin{theo}\label{theo:main}
$\pEIS$ is fixed-parameter tractable.
\end{theo}

\proof We define a computable function $D_0: \mathbb N\to \mathbb N$
by
\begin{align}\label{eq:mainD}
D_0(k) & := 2\cdot \left((p_k-1)^{p_k+1}+(p_k-1)^2\right)
 + 2^{2\cdot ((k-1)^2+1)}.
\end{align}
Note $2^{2\cdot ((k-1)^2+1)}> \mathscr R_{(k-1)^2+1}$. Then let
$\mathbb A_{D_0}$ be the algorithm as stated in Lemma~\ref{lem:main}
for the function $D_0$.

\medskip
Let $(G,k)$ with $G=(V,E)$ be an instance of $\pEIS$. First, we
remove all the isolated vertices in $G$. For simplicity, the
resulting graph is denoted by $G$ again. Then, we simulate the
algorithm $\mathbb A_{D_0}$ on $(G,k)$. If the result is either (S1)
or (S2) in Lemma~\ref{lem:main}, we already get the correct answer.
Otherwise, $\mathbb A_{D_0}$ outputs three subsets
$V_1,V_2,B\subseteq V$ satisfying (S3.1) and (S3.2).

If $G[V_1]$ and $G[V_2]$ are both $D_0(k)$-degree-extreme, then
$(G,V_1, V_2, B)$ is a $(D_0(k), |B|)$-bridge with $|B|$ bounded by
an appropriate computable function of $k$. The fixed-parameter
tractability of whether $G$ contains a $k$-edge induced subgraph
follows from Proposition~\ref{prop:mcbs}. Otherwise, either $G[V_1]$
or $G[V_2]$ is not $D_0(k)$-degree-extreme.

\medskip
We assume that $G[V_1]$ is not $D_0(k)$-degree-extreme. (The case
for $G[V_2]$ is symmetric.) Then we simulate the algorithm $\mathbb
A_{D_0}$ on $(G[V_1],k)$. Observe that the result cannot be (S1). If
the output is (S2), since $G[V_1]$ is an induced subgraph of $G$, we
conclude that $G$ has an induced subgraph of exactly $k$ edges.

\medskip
Now we are left with case (S3). In particular, there are subsets
$V_{11}, V_{12}, B_1\subseteq V_1$ such that the corresponding
properties of (S3.1) and (S3.2) are satisfied. Let
\[
U_1 := V_{11}\setminus (B\cup B_1), \quad
U_2 := V_{12}\setminus (B\cup B_1), \quad \text{and}\ \
U_3 := V_{2}\setminus (B\cup B_1).
\]
Observe that in $G$ if we remove the vertex set $B$, then there is
no edge left between $V_1$ and $V_2$. Similarly, if we remove the
vertex set $B_1$, every edge between $V_{11}$ and $V_{12}$ is
destroyed. Thus, by (S3.2), in the original graph $G$, there is no
edge between each pair of $U_1$, $U_2$ and $U_3$. Moreover by (S3.1)
and (S3.2) for every $i\in [3]$
\[
|U_i|> D_0(k)- 2\cdot \left((p_k-1)^{p_k+1}+(p_k-1)^2\right)
= 2^{2\cdot ((k-1)^2+1)}>\mathscr R_{(k-1)^2+1},
\]
where the equality is by~\eqref{eq:mainD}.

We use Ramsey's Theorem again. If there is an independent set of
size $(k-1)^2+1$ in one of the $U_1$, $U_2$ and $U_3$, as $G$ has no
isolated vertex, then $G$ contains a $k$-edge induced subgraph by
Lemma~\ref{lem:is}. Otherwise every $U_i$ contains a clique of size
$(k-1)^2+1 \ge k$. As we have seen that there is no edge between
$U_1$, $U_2$ and $U_3$ in $G$, Lemma~\ref{lem:3cliques} implies that
$G$ contains an induced subgraph of exactly $k$ edges. \proofend

\begin{rem}\label{rem:fpttime}
We mentioned in the Introduction that the running time of our
\fpt-algorithm in terms of $k$ is \emph{triple exponential} at
least. To see this, recall the function $D_0$ as defined
in~\eqref{eq:mainD} is of the order $2^{2^{\Theta(k)}}$. This gives
the quadruple exponential lower bound for the algorithm $\mathbb
A_{D_0}$ by Remark~\ref{rem:mcdestime}. So the same lower bound
applies to our algorithm for \pEIS.
\end{rem}

\section{Counting $k$-edge induced subgraphs}\label{sec:count}

In this section we study two counting versions of $\pEIS$. Of
course, the most natural version is:
\npprob{\pCEIS}{A graph $G$ and $k\in \mathbb N$}{$k$}{Compute the
number of $k$-edge induced subgraphs in $G$.}

In general, a parameterized counting problem is a pair $(F,\kappa)$,
where $F:\Sigma^*\to \mathbb N$ and $\kappa$ is a parameterization.
$(F,\kappa)$ is fixed-parameter tractable if $F$ can be computed by
an \fpt-algorithm with respect to $\kappa$. For more background of
parameterized counting complexity, the reader is referred
to~\cite{flugro04,mcc}.

\medskip
In fact, the hardness of $\pCEIS$ is rather easy to show. We observe
that the vertex set of every induced subgraph \emph{without any
edge} is an independent set, and vice versa. Hence the \emph{first
slice} of $\pCEIS$, i.e., counting the number of $0$-edge induced
subgraphs is exactly the classical problem:
\nprob{$\#\textsc{Independent-Set}$}{A graph $G$}{Compute the number
of independent sets in $G$.}
Recall that $\#\textsc{Independent-Set}$ is
$\#\Pt$-hard~\cite{val,probal}. Hence:

\begin{theo}
Assume $\#\Pt \ne \Pt$. Then $\pCEIS$ is not fixed-parameter
tractable.
\end{theo}

One might attribute the above hardness result to the fact that we
allow induced subgraphs to have isolated vertices. Note these
isolated vertices play no role in the decision problem $\pEIS$.
Therefore, it also makes sense to consider:
\npprob{$\pCEIS^*$}{A graph $G$ and $k\in \mathbb N$}{$k$}{Compute
the number of $k$-edge induced subgraphs \emph{without isolated
vertices} in $G$.}

Then we show:
\begin{theo}\label{theo:pCEISs}
$\pCEIS^*$ is hard for $\#\W 1$.
\end{theo}

Here, $\#\W 1$ is the counting version of the parameterized class
$\W 1$. One standard complete problem of $\#\W 1$ is:
\npprob{$p\textsc{-\#Independent-Set}$}{A graph $G$ and $k\in
\mathbb N$}{$k$}{Compute the number of independent sets of size $k$
in $G$.}
To prove the $\#\W 1$-hardness, we need an appropriate notion of
reduction. Let $(F,\kappa)$ and $(F',\kappa')$ be two parameterized
counting problems. An \emph{\fpt\ Turing reduction} from $(F,
\kappa)$ to $(F', \kappa')$ is an algorithm $\mathbb A$ with an
oracle to $F'$ which satisfies the following conditions:
\begin{itemize}
\item $\mathbb A$ computes the function $F$ in \fpt-time (with
    respect to $\kappa$).

\item There is a computable function $g:\mathbb N\to \mathbb N$
    such that for all oracle queries ``$F'(y)=?$'' posed by
    $\mathbb A$ on input $x$ we have $\kappa'(y) \le
    g(\kappa(x))$.
\end{itemize}
It is easy to verify that if $(F,\kappa)$ is $\#\W 1$-hard and there
is an \fpt\ Turing reduction from $(F, \kappa)$ to $(F', \kappa')$,
then $(F',\kappa')$ is $\#\W 1$-hard.

\medskip
\noindent \textit{Proof of Theorem~\ref{theo:pCEISs}:} We give an
\fpt\ Turing reduction from $p\textsc{-\#Independent-Set}$ to
$p\textsc{-\#Edge}$\-$\textsc{Induced-Subgraph}^*$. To simplify the
presentation, let us call an induced subgraph without isolated
vertices \emph{nice}.

\medskip
Let $(G,k)$ be an instance of $p\textsc{-\#Independent-Set}$. For
each $i\in [k]$ we define $V_{2\cdot i-1} := \big\{(v,i) \bigmid
v\in V(G)\big\}$. Moreover, for $i\in [k-1]$ let $V_{2\cdot i} :=
\{e_i\}$, where all $e_i$'s are new vertices not in $V(G)$. Then we
define a new graph $H$ with
\begin{align*}
V(H) & := \bigcup_{i\in [2\cdot k-1]} V_i \\
E(H) & := \bigcup_{i\in [k]} \big\{\{(u,i), (v,i)\}\bigmid
 \text{$u,v\in V(G)$ with $u\ne v$}\big\} \\
 & \quad \cup \bigcup_{1\le i<j \le k} \big\{\{(u,i), (v,j)\}
 \bigmid \text{$u=v$ or $\{u,v\}\in E$}\big\} \\
 & \quad \cup \bigcup_{i\in [k-1]}
 \big\{\{(v,j), e_i\} \bigmid \text{$v\in V(G)$ and ($j=i$ or $j=i+1$)}
 \big\}.
\end{align*}
For each $i\in [2\cdot k-1]$ we call $V_i$ a \emph{block} of $G$.
Observe that each odd block is a clique of size $|V(G)|$ and each
even block a singleton set.

Let $\{v_1,\ldots, v_k\}\subseteq V(G)$ be an independent set of
size $k$ in $G$. Clearly
\[
G\Big[\big\{(v_i,i) \bigmid i\in [k]\big\} \cup
\big\{e_i \bigmid i\in [k-1]\big\}\Big]
\]
is a $(2\cdot k-2)$-edge nice induced subgraph of $G$. The crucial
observation is that the following converse is also true.

\medskip
\noindent \textit{Claim.} Let $H'$ be a nice induced subgraph of $H$
containing exactly $2\cdot k-2$ edges. If $V(H') \cap V_i \ne
\emptyset$ for every $i\in [2\cdot k-1]$, i.e., $H'$ intersects all
blocks $V_i$'s, then
\[
\big\{v\in V \mid \text{for some $i\in [k]$ we have
$(v,2\cdot i-1)\in V(H')$}\big\}
\]
is an independent set in $G$ of size $k$.

\medskip
\noindent \textit{Proof of the claim.} First we show that
$|V(H')\cap V_i|=1$ for all $i\in [2\cdot k-1]$. This is obviously
true for even $i$'s, i.e., $H'$ contains all $e_i$'s. As $e_i$ is
adjacent to every vertex in the blocks $V_{2\cdot i-1}$ and
$V_{2\cdot i+1}$, if $H'$ contains two vertices in one odd block,
then $H'$ would have more than $2\cdot k -2$ edges, a contradiction.

Next for every $i\in [k]$ let $v_i$ be the vertex in $G$ such that
$V(H')\cap V_{2\cdot i-1} = \big\{(v_i, 2\cdot i-1)\big\}$. At this
point, we already know that $H'$ contains the following $2\cdot k-2$
edges
\begin{equation}\label{eq:2k1}
\{v_1,e_1\}, \{e_1, v_2\}, \ldots, \{v_{k-1}, e_k\},
\{e_k, v_k\}.
\end{equation}
We prove that $\{v_1,\ldots, v_k\}$ is an independent set in $G$ of
size $k$. Otherwise for some $1\le i<j \le k$ we have $v_i= v_j$ or
$\{v_i,v_j\}\in E(G)$. Then $H'$ would contain a further edge
$\big\{(v_i,2\cdot i-1), (v_j, 2\cdot j -1)\big\}$ and hence have
more than $2\cdot k-2$ edges by~\eqref{eq:2k1}. \hfill$\dashv$

\medskip
It follows that
\begin{align}\notag
 \text{\big(the} &\ \text{number of independent sets of size $k$ in $G$\big)}
 \cdot \ k! \\\label{eq:is}
 & = \text{the number of $(2\cdot k-2)$-edge nice induced subgraphs in $H$
which intersect every $V_i$}.
\end{align}
Thus our goal is to compute the right hand side of \eqref{eq:is}
using $\pCEIS$ as an oracle. To that end for every $X\subseteq
[2\cdot k -1]$ we let
\[
H_X := H\left[\bigcup_{i\in X} V_i\right]
\]
and
\begin{align*}
s_X & := \text{the number of $(2\cdot k-2)$-edge nice induced subgraphs in $H_X$}, \\
t_X & := \text{the number of $(2\cdot k-2)$-edge nice induced subgraphs in $H_X$} \\
& \hspace{6cm} \text{which intersect $V_i$ for every $i\in X$}.
\end{align*}
Therefore, the right hand side of~\eqref{eq:is} is exactly
$t_{[2\cdot k-1]}$.

Note every $s_X$ can be computed by an oracle query to $\pCEIS$ on
the instance $(H_X, 2\cdot k-2)$. Moreover it is easy to see
\begin{equation*}
t_X = s_X- \sum_{Y\subsetneq X} t_Y.
\end{equation*}
Hence, by simple dynamic programming using $\pCEIS$ as an oracle, we
can compute every $t_X$ in \fpt\ time. \proofend

\medskip
\subsection*{Acknowledgement} We thank Leizhen Cai for bringing the
problem \pEIS\ to our attention, and J\"org Flum for comments on
earlier versions of this paper.


\bibliographystyle{plain}
\bibliography{kedge}

\newpage
\section*{Appendix}
For the reader not familiar with~\cite{frigro} we give a detailed
proof of Theorem~\ref{theo:mcltwp0}. Our presentation closely
follows that of~\cite[Section~12.2]{flugro}. Overall we will reduce
$\pMCLTWFO g$ to a generalization of the parameterized independent
set problem.

\begin{defn}\label{def:lrscatter}
Let $G=(V,E)$ be a graph and $\ell, r\in \mathbb N$. A set
$S\subseteq V$ is \emph{$(\ell,r)$-scattered} if there exist
$v_1,\ldots, v_{\ell} \in S$ such that for every $1\le i< j\le \ell$
we have $d(v_i,v_j)> r$.
\end{defn}

\begin{prop}\label{prop:scatterltw}
Let $g:\mathbb N\times \mathbb N\to \mathbb N$ be a computable
function. Then the following parameterized problem is
fixed-parameter tractable.
\npprob{$\pSCLTW g$}{A graph $G=(V,E)$, $S\subseteq V$ and $p,\ell,
r\in \mathbb N$ such that $G$ has local tree-width bounded by $g$
with respect to $p$}{$p+\ell+r$}{Decide whether $S$ is
$(\ell,r)$-scattered.}
\end{prop}

To prove this proposition we need another simple combinatorial
result (for a proof see, e.g., \cite[Lemma~12.12]{flugro}).

\begin{lem}\label{lem:dsdiameter}
Let $G=(V,E)$ be a connected graph and $S\subseteq V$ a dominating
set\footnotemark\ in $G$. Then $d(u,v) \le 3\cdot |S|- 1$ for every
$u,v \in V$. That is, the diameter of $G$ is bounded by $3\cdot |S|-
1$. \footnotetext{Recall, $S\subseteq V(G)$ is a dominating set if
for every $u\in V(G)$ either $u\in S$ or there is a vertex $v\in S$
with $\{u,v\}\in E(G)$.}
\end{lem}

\medskip
\noindent \textit{Proof of Proposition~\ref{prop:scatterltw}:} By
Courcelle's Theorem~\cite{cou} it is easy to see that the problem
\npprob{$\pSCTW$}{A graph $G=(V,E)$, $S\subseteq V$ and $\ell, r\in
\mathbb N$}{$\tw(G)+\ell+r$}{Decide whether $S$ is
$(\ell,r)$-scattered.}
is fixed-parameter tractable. So our goal is to give an
\fpt-reduction from $\pSCLTW g$ to $\pSCTW$.

\medskip

First, using a simple greedy algorithm, we can compute in linear
time a \emph{maximal} set $T\subseteq S$ such that for every
distinct $u,v\in T$ we have $d^{G}(u,v)>r$. If $|T| \ge \ell$, then
we are done. Otherwise
\begin{equation}\label{eq:Tl}
|T|< \ell.
\end{equation}

\noindent \textit{Claim 1.} $S\subseteq N^G_{r}(T)$ \Big($:=
\big\{v\in V\bigmid \text{$d^G(u,v)\le r$ for some vertex $u\in
T$}\big\}$\Big).

\medskip
\noindent \textit{Proof of the claim.} Otherwise let $v\in
S\setminus N^G_{r}(T)$. Thus $d^G(v,u)> r$ for every $u\in T$. This
contradicts the maximality of $T$. \hfill$\dashv$

\bigskip \noindent \textit{Claim 2.} $S$ is $(\ell,r)$-scattered in
$G$ if and only if $S$ is $(\ell,r)$-scattered in $\mathcal
N^G_{2\cdot r}(T)$ \Big( $:= G\big[N^G_{2\cdot r}(T)\big]$\Big).

\medskip
\noindent \textit{Proof of the claim.} The direction from left to
right is trivial. So let us assume that $S$ is $(\ell,r)$-scattered
in $\mathcal N^G_{2\cdot r}(T)$. In particular, there exist
$v_1,\ldots, v_{\ell}\in S$ such that
\begin{equation}\label{eq:dvivj}
d^{\mathcal N^G_{2\cdot r}(T)}(v_i,v_j)> r
\end{equation}
for every $1\le i< j\le \ell$. Towards a contradiction assume that
there exist some $i,j \in \mathbb N$ with $1\le i<j \le \ell$ and
$d^{G}(v_i,v_j)\le r$. Note every vertex $u$ in a shortest path
between $v_i$ and $v_j$ satisfies $d^{G}(u,v_i)\le r$, and hence,
$u\in N^G_r(S) \big(:= \big\{v\in V\bigmid \text{$d^G(u,v)\le r$ for
some vertex $u\in S$} \big\}\big)$. Then by Claim~1, $u \in
N^G_{2\cdot r}(T)$. As a consequence $d^{\mathcal N^G_{2\cdot
r}(T)}(v_i,v_j)\le r$, which contradicts~\eqref{eq:dvivj}.
\hfill$\dashv$

\medskip
Claim~2 shows that the mapping
\[
R(G,S,p,\ell,r) := \left(\mathcal N^G_{2\cdot r}(T),S,\ell,r\right)
\]
is a correct reduction from $\pSCLTW g$ to $\pSCTW$. It remains to
show $R$ is an \fpt-reduction. To that end, we need to bound
$\tw\left(\mathcal N^G_{2\cdot r}(T)\right)+\ell+r$ in terms of
$p+\ell+r$.

\medskip
\noindent \textit{Claim 3.} $\tw\left(\mathcal N^G_{2\cdot
r}(T)\right) \le g(2\cdot r\cdot (3\cdot \ell-4),p)$.

\medskip
\noindent \textit{Proof of the claim.} Let $H$ be a graph with
\[
V(H) := N^G_{2\cdot r}(T) \ \text{and}\
E(H) := \big\{\{u,v\} \mid \text{$u,v\in V(H)$, $u\ne v$ and
$d^G(u,v) \le 2\cdot r$}\big\}.
\]
It is then easy to verify that $T$ is a dominating set in $H$. Hence
by Lemma~\ref{lem:dsdiameter}, every connected component of $H$ has
diameter at most $3\cdot |T|-1 \le 3\cdot \ell-4$ by~\eqref{eq:Tl}.
It follows that every connected component $C$\ of $\mathcal
N^G_{2\cdot r}(T)$ has diameter at most $2\cdot r\cdot (3\cdot
\ell-4)$. This implies that $C= N^G_{2\cdot r\cdot (3\cdot
\ell-4)}(v)$ for every $v\in C$. Recall that $G$ has local
tree-width bounded by $g$ with respect to $p$. Hence,¡¡
\[
\tw\left(\mathcal N^G_{2\cdot r}(T)\right)
 \le g(2\cdot r\cdot (3\cdot \ell-4),p) \tag*{$\dashv$}
\]

This finishes the proof. \proofend

Now we recall Gaifman's Theorem~\cite{gai}.

\begin{lem}
Let $\tau$ be a vocabulary and $r\in \mathbb N$. Then there is an
\FO-formula $\delta_r(x,y)$ such that for all $\tau$-structure
$\mathcal A$ and all elements $a,b\in A$ we have $d^{G(\mathcal
A)}(a,b) \le r$ if an only if $\mathcal A\models \delta_r(a,b)$.

For simplicity we will write $d(x,y)\le r$ and $d(x,y)> r$ instead
of $\delta_r(x,y)$ and $\neg \delta_r(x,y)$, respectively.
\end{lem}

An \FO\ $\tau$-formula $\psi(x)$ is \emph{$r$-local} if for all
$\tau$-structure $\mathcal A$ and $a\in A$:
\begin{eqnarray*}
\mathcal A \models \psi(a) & \iff &
\mathcal N^{\mathcal A}_r(a) \models \psi(a).
\end{eqnarray*}

\begin{theo}[\bf Gaifman's Theorem]\label{theo:gaifman}
Every \FO-sentence $\varphi$ is equivalent to a Boolean combination
of sentences of the form
\[
\exists x_1\ldots \exists x_{\ell}\left(\bigwedge_{1\le i< j\le\ell}
 d(x_i,x_j)>2\cdot r
\wedge \bigwedge_{i\in [\ell]} \psi(x_i)\right).
\]
with $\ell,r \in \mathbb N^+$. Moreover, such a Boolean combination
can be computed from $\varphi$.
\end{theo}

Now we have all the tools for proving Theorem~\ref{theo:mcltwp0}
which for the reader's convenience we repeat as below:

\begin{theo}\label{theo:mcltwp}
For every computable function $g: \mathbb N\times \mathbb N \to
\mathbb N$ the problem $\pMCLTWFO g$ is fixed-parameter tractable.
\end{theo}

\proof Let $(\mathcal A, p, \varphi)$ be an instance of \pMCLTWFO g.
It is easy to see that, by Gaifman's Theorem, we can assume without
loss of generality that for some $\ell,r\in \mathbb N$ and $r$-local
\FO-formula $\psi$
\[
\varphi= \exists x_1\ldots \exists x_{\ell}\left(\bigwedge_{1\le i< j\le\ell}
 d(x_i,x_j)>2\cdot r\wedge \bigwedge_{i\in [\ell]} \psi(x_i)\right).
\]

\medskip
Let $G=(V,E)$ be a graph with $V := A$ and $E:= \big\{ \{a,b\}
\bigmid \text{$a,b\in A$ and $d^{G(\mathcal A)}(a,b)=1$}\big\}$.
That is, $G$ is Gaifman's graph of $\mathcal A$. Moreover, let $S:=
\big\{a\in A\bigmid \mathcal A\models \psi(a) \big\}$. By the
$r$-locality of $\psi$ we have $S= \left\{a\in A\bigmid \mathcal
N^{\mathcal A}_r(a)\models \psi(a) \right\}$. Since
$\tw\left(\mathcal N^{\mathcal A}_r(a)\right) \le g(r,p)$, we can
compute the set $S$ in \fpt\ time, again by Courcelle's Theorem.

It is now easy to verify that $\mathcal A\models \varphi$ if and
only if $S$ is $(\ell, r)$-scattered in $G$, i.e.,
\begin{eqnarray*}
(\mathcal A, p, \varphi)\in \pMCLTWFO g & \iff &
(G,S,p,\ell)\in \pSCLTW g.
\end{eqnarray*}
Now the result follows from Proposition~\ref{prop:scatterltw}.
\proofend

\end{document}